\documentclass[12pt,a4paper,twoside]{article}
\usepackage{amsmath}
\usepackage{amssymb,latexsym}
\usepackage{graphicx}

\def\tr{\operatorname{tr\,}}

	\def\Im{{\rm Im\,}}	  
		  
		  \def\deg{{\rm deg\,}}
\def\det{{\rm det\,}}

    \def\cC{{\cal C}}
    \def\cF{{\cal F}}

    \def\cO{{\cal O}}

  \def\cW{{\cal W}}

\def\id{\protect{{1 \kern-.28em {\rm l}}}}
\newcommand{\vev}[1]{\left\langle{#1}\right\rangle}

\def\None{{\cal N}\!=\!1}	\def\Ntwo{{\cal N}\!=\!2}

 \def\Nt{\widetilde{N}} \def\Qt{\widetilde{Q}}
\def\Rt{\widetilde{R}}  \def\Ut{\widetilde{U}}
 \def\Wt{\widetilde{W}} \def\st{\widetilde{s}}
 
\newcommand{\wtil}[1]{\widetilde{#1\,}}

  \def\hb{{\widehat B}}

\newcommand{\be}{\begin{equation}} 	\newcommand{\ee}{\end{equation}}
\newcommand{\bea}{\begin{eqnarray}}	\newcommand{\eea}{\end{eqnarray}}
\newcommand{\beann}{\begin{eqnarray*}}	\newcommand{\eeann}{\end{eqnarray*}}
\newcommand{\bfig}{\begin{figure}}	\newcommand{\efig}{\end{figure}}
\newcommand{\ba}{\begin{array}}		\newcommand{\ea}{\end{array}}
\newcommand{\bcen}{\begin{center}}	\newcommand{\ecen}{\end{center}}
\newcommand{\nn}{\nonumber}

\newtheorem{Proposition}{Proposition}[section]

\newtheorem{Theorem}{Theorem}[section]
\newtheorem{Lemma}{Lemma}[section]
\newtheorem{Corrolary}{Corrolary}[section]

\newcommand{\bp}{\begin{Proposition}}	\newcommand{\ep}{\end{Proposition}}
\newcommand{\bt}{\begin{Theorem}}	\newcommand{\et}{\end{Theorem}}
\newcommand{\bl}{\begin{Lemma}}		\newcommand{\el}{\end{Lemma}}
\newcommand{\bc}{\begin{Corrolary}}	\newcommand{\ec}{\end{Corrolary}}

\def\erf#1{(\ref{#1})}

\begin{document}

\begin{titlepage}

\begin{flushright}
IFT-UAM/CSIC-05-25 \\
\texttt{hep-th/0505119}
\end{flushright}

\vspace{2cm}

\bcen
{\LARGE\bfseries Dual Interpretations of Seiberg--Witten \\[1ex] and Dijkgraaf--Vafa curves}\\[3em]
{\bfseries K. Landsteiner\footnote{\tt Karl.Landsteiner@uam.es} and S. Montero\footnote{\tt Sergio.Montero@uam.es}}\\[2em]
{\it Instituto de F\'\i sica Te\'orica, \\
C-XVI Universidad Aut\'onoma de Madrid, \\
28049 Madrid, Spain}
\ecen

\vspace{3cm}

\abstract{We give dual interpretations of Seiberg--Witten and Dijkgraaf--Vafa (or matrix model) curves in $\None$ supersymmetric $U(N)$ gauge theory. This duality interchanges the rank of the gauge group with the degree of the superpotential; moreover, the constraint of having at most log-normalizable deformations of the geometry is mapped to a constraint in the number of flavors $N_f < N$ in the dual theory.}
\end{titlepage}

\setcounter{footnote}{0}
\setcounter{page}{1}
\section{Introduction} 

Supersymmetric gauge theories are of great interest not only as possible extensions of the standard model of particle physics, but also as a theoretical laboratory to develop and test ideas about the strong coupling dynamics of gauge theories in general. Holomorphicity arguments allow to obtain exact non-perturbative results. In $\None$ supersymmetric gauge theories, the exactly computable holomorphic quantities include the superpotential. Relying on arguments from string theory Dijkgraaf and Vafa argued \cite{DVpapers} that the exact superpotential can be computed from the partition function of a simple matrix model; the action of it being given by the tree level superpotential of the gauge theory. The loop equations of the matrix model gives rise to an algebraic Riemann surface in the large $N$ limit and the partition function can be calculated from a set of contour integrals on this Riemann surface. The superpotential can then in turn be computed from the partition function of the matrix model.  

Using the special properties of chiral operators in supersymmetric theories, Cachazo {\it et al.} showed in \cite{CIV} that the matrix model results can be obtained directly in gauge theory through a generalization of the Konishi anomaly equations \cite{cpsk}. The Konishi anomaly relations give rise to the same Riemann surface as the matrix model loop equations and so, the superpotential of the gauge theory can be calculated in terms of the same contour integrals that determine the matrix model partition function. 

One of the most studied examples in this context is $\None$ $U(N)$ gauge theory with a chiral matter multiplet $\phi$ in the adjoint representation, $N_f$ chiral multiplets in the fundamental and antifundamental representation and a tree level superpotential of the form
\begin{equation}
\label{Wtree}
W_{\mathrm{tree}} = W(\phi) + \sum_{f=1}^{N_f} \widetilde Q^f (\phi + m_f) Q_f~~~,~~~ W(\phi)=\sum_{k=0}^m \frac{g_k}{k+1} \tr(\phi^{k+1}) 
\end{equation} 
Classically this model has a set of discrete vacua characterized by the eigenvalues $a_i$ of the vacuum expectation value of $\phi$. Each eigenvalue can be $N_i$ times degenerate, the gauge group is broken to $\prod_{i=1}^m U(N_i)$ and $\sum_{i=0}^m N_i = N$. The eigenvalues $a_i$ are the solutions of $W'(z)=0$ and the vacuum expectation values of the fundamentals vanish. In addition there are Higgs vacua with $\langle  \Qt^f Q_f\rangle \neq 0$ and a corresponding eigenvalue of $\phi$, which we call $z_f = -m_f$. In the Higgsed vacua the rank of the gauge group is reduced such that $\sum_{i=1}^m N_i < N$. 

Although Higgs and Coulomb vacua are clearly distinguished phases classically, it is well known that in theories with matter transforming faithfully under the center of the gauge group there is no phase distinction between the two \cite{FSBR}. In supersymmetric gauge theories this is also well-established in $\Ntwo$ \cite{SWmatter} and in the context of generalized Konishi anomalies in \cite{andi} and in particular in \cite{CSWmatter}.

The generalized Konishi anomaly relations determine relations for the vacuum expectation values of the generating functions
\begin{equation}
T(v) =  \left\langle \frac{1}{v-\phi}\right\rangle ~~~~,~~~
R(v) =  \left\langle -\frac{1}{32\pi^2}\frac{ \cW^\alpha \cW_\alpha}{v-\phi}\right\rangle\,.
\end{equation} 
Introducing $y(v) = 2R(v) + W'(v)$ the Konishi anomaly relation for $R$ takes the form of a hyperelliptic Riemann surface
\begin{equation}\label{Sigma}
y^2 = (W'(v))^2 - f_{m-1}(v)\,,
\end{equation} 
where $f_{m-1}$ is a polynomial of degree $m-1$ and its coefficients determine the gaugino condensates $S_i$ in the $i$-th factor group.
 
The solution for $T(v)$ is
\begin{equation}
\label{Tsol}
T(v) = \sum_{f=1}^{N_f} \left( \frac{1}{2(v+m_f)} + \frac{(1-2r_f)y(q_f)}{2 y(v) (v+m_f)}\right) + \frac{c(v)}{y(v)} \,, 
\end{equation}  
where $q_f$ is the point on the Riemann surface \erf{Sigma} lying above $v=-m_f$ on the upper sheet \footnote{The Riemann surface \erf{Sigma} is a two-sheeted cover of the $v$-plane. The sheet on which $y(v) \approx S/v$ for large values of $v$ is called upper sheet.}, $r_f$ is one if the $f$-th flavor gives rise to higgsing, otherwise $r_f=0$ and $c(v)$ is a polynomial of degree $m-1$ whose coefficients determine the ranks $N_i$ of the factor groups. The gaugino condensates and the ranks of the factor groups are determined by the compact period integrals on $\Sigma$ as
\begin{equation}
\label{periods}
S_i = \oint_{A_i} R(v) \, dv \qquad,\qquad N_i = \oint_{A_i} T(v) \, dv ~.
\end{equation} 
These expressions allow to compute the effective superpotential off-shell as a holomorphic function of the gaugino bilinears $S_i$. The on-shell superpotential is obtained then by minimizing with respect to the $S_i$. On-shell, all the compact periods of $T$ are integer valued \cite{CSWmatter}. This demands that the on-shell $T$ is the derivative of the logarithm of a meromorphic function on \erf{Sigma}.
It can be written as
\begin{equation}
T(v) = \frac{\partial}{\partial v}\log \left( P(v) + \sqrt{P^2 - 4 \Lambda^{2N-N_f} \prod_{f=1}^{N_f} (v+m_f)}\right) \,,
\end{equation} 
where $P(v)$ is a polynomial of degree $N$. The condition that $T(v)$ is a meromorphic differential on $\Sigma$
implies the factorization 
\begin{eqnarray}\label{factorization}
y_{\rm SW}^2 = P_N^2(v) - 4\Lambda^{2N-N_f} \prod_{f=1}^{N_f} (v+m_f) &=& Q_{2n}(v) S_{N-n}(v)^2 \,,\\
y_{m.m.}^2 = (W'_m(v))^2 - f_{m-1}(v) &=& Q_{2n}(v) H_{m-n}(v)^2\,,
\end{eqnarray} 
where $F$, $H$ and $Q$ are polynomials of the degree indicated as an index. It is furthermore obvious that the first equation is the Seiberg--Witten \cite{SWnomatter} curve of the $\Ntwo$ gauge theory that is obtained by setting the tree level couplings $g_k$ to zero. From the standpoint of Seiberg--Witten theory the factorization reflects the fact that addition of a superpotential in $\phi$ localizes the Coulomb branch of the $\Ntwo$ moduli space to the subloci where $n$ monopoles (or dyons) become massless. These are precisely the loci where the Seiberg--Witten curve develops $n$ double points. A particular solution to this factorization condition \erf{factorization} represents a choice of vacuum in the gauge theory. Moreover, the solutions to the factorization conditions include vacua on the pseudo-confining as well as the pseudo-higgs branch, and both branches can be connected by varying the masses such that $q_f$, the image point of $m_f$ on \erf{Sigma} crosses a branch cut \cite{CSWmatter,AFOS}.

These factorization conditions have been given a beautiful interpretation in \cite{deboer}. The gauge theory is realized on a single M-theory fivebrane whose worldvolume is partially wrapped on a Riemann surface embedded in a three dimensional complex space. The Riemann surface is described by two algebraic equations. Each of them can be seen as a projection of the Riemann surface onto a two dimensional subspace.  To represent the ground state of the gauge theory the two Riemann surfaces have to be proportional to each other, which means that they have to factorize onto a common reduced Riemann surface just as in \erf{factorization}, i.e. the Riemann surface on which the \mbox{M-theory} fivebrane is wrapped is given precisely by equations \erf{factorization}. We point out that there are two different $\None$ gauge theories with different rank and superpotential who produce the same projections. This observation is the starting point for our argument. We will see that the same set of equations and factorization conditions can be obtained from two different gauge theories. What in one gauge theory represents the Seiberg--Witten curve is interpreted as the matrix model curve in the other gauge theory and vice versa. Thus we find pairs of ``dual" gauge theories which in some sense share the same ground state. A somewhat surprising feature of this ``duality" is that the rank of the first gauge group determines the degree of the tree level superpotential of the other gauge theory. 

\noindent This paper is organized as follows. In section \ref{sec:M5brane} we review first the M-theory fivebrane construction of $\None$ supersymmetric gauge theories and show how the same curve emerges from two different gauge theories. 

In section  \ref{sec:dual-spots} we compute the exact superpotential of the ``dual" gauge theories. In order to do so we first prove that the exact superpotential as computed from the geometry of the Riemann surface coincides with the low energy field theoretical superpotential. Here we extend the known proofs in the literature to the case of arbitrary high degree superpotential and matter content. We then compute explicitly the superpotentials in two simple cases.

Section \ref{sec:conclusions} discusses the results obtained and offers some conclusions.

\setcounter{equation}{0}
\section{The M-theory fivebrane}
\label{sec:M5brane}
One of the most useful tools to model non-perturbative dynamics of supersymmetric gauge theories is the M-theory fivebrane. In particular, four dimensional gauge theories can be constructed on the worldvolume of configurations of intersecting D- and NS-branes in type IIA string theory in the weak coupling limit. The strong coupling dynamics is modelled by going to the strong coupling  M-theory limit of IIA string theory. 

The best known and studied example is $SU(N)$ supersymmetric gauge theory, \mbox{realized} on the worldvolume of D4-branes that are spanned between two NS5-branes. Two different situations can be distinguished: one with parallel fivebranes and resulting in $\Ntwo$, the other one with non-parallel fivebranes resulting in $\None$. In the following we will review the construction in detail and also recall the beautiful interpretation as a unification of DV (Dijkgraaf--Vafa) and SW (Seiberg--Witten) curves as pointed out in \cite{deboer}.

Let us start in type IIA string theory. The NS-fivebranes will span the directions $x^0,\ldots,x^3,x^4,x^5$ and are separated by some distance, say $L$, in the $x^6$ direction. Otherwise they are parallel. In between these fivebranes we span $N$ D4-branes along the directions $x^0,\ldots,x^3,x^6$. In the $x^6$ direction the D4-branes terminate on the fivebranes. This brane construction engineers an $\Ntwo$ supersymmetric gauge theory with gauge group $SU(N)$ on the non-compact, $3+1$ dimensional part of the worldvolume of the D4-branes \cite{edi}. So far we have constructed pure gauge theory. To include also matter (hyper)multiplets in the fundamental representation of $SU(N)$ there are two ways to achieve this. One way is to use D6-branes spanning the $x^0,\ldots,x^3, x^7,x^8,x^9$ directions and located in between the NS-branes in the $x^6$ direction. An alternative is to consider semi-infinite D4-branes ending on only one of the NS-branes, e.g.  D4-branes spanning all of $x^6$ (and $x^0,\ldots,x^3$) to the right of the rightmost NS-brane. Both possibilities are equivalent through the so called Hanany--Witten transition \cite{HW}. We will however focus in this paper on the realization of fundamental matter through D4-branes only.

This type IIA brane configuration can be lifted to M-theory and represents then the strong-coupling dynamics of the gauge theory. In M-theory, the D4-branes are lifted to M-theory fivebranes wrapped on the additional circle and what appears as intersecting branes in IIA string theory is lifted to a single M-theory fivebrane wrapping a complex curve. We introduce complex coordinates $v = x^4 + i x^5$  and $t = \exp(-(x^6+i x^{10})/R)$, where $x^{10}$ parameterizes the eleventh dimension of M-theory, it is supposed to be compact with identification $x^{10} \equiv x^{10} + 2\pi R$. Type IIA string theory is recovered in the limit $R\rightarrow 0$. The M-theory fivebrane is partially wrapped on the Riemann surface 
\begin{equation}\label{swcurve1st}
t^2 - 2 P_N(v) t + 4 \Lambda^{2N-N_f} \prod_{f=1}^{N_f} (v+m_f) =0\,,
\end{equation} 
Here $P_N(v)$ is a polynomial of degree $N$ in $v$. In the classical or type IIA limit the zeroes of $P_N(v)$ correspond to the locations of the D4-branes in the $v$-plane. Similarly $m_f$ are the positions of the semi-infinite fourbranes to the right and represent the masses of the fundamental hypermultiplets in the gauge theory. Of course, the Riemann surface \erf{swcurve1st} is the Seiberg--Witten curve of the corresponding $\Ntwo$ supersymmetric gauge theory with $N_f$ hypermultiplets in the fundamental representation. An additional ingredient in Seiberg--Witten theory is the differential $\lambda_{\rm SW} = v \frac{dt}{t}$. Its period integrals around the cuts of \erf{swcurve1st} define special coordinates $a_i$ and the prepotential $\cF(a_i)$ as
\begin{equation}
a_i = \frac{1}{2\pi i}\oint_{A_i} \lambda_{\rm SW} \qquad,\qquad \frac{\partial \cF}{\partial a_i} = \frac{1}{2\pi i}\oint_{B_i} \lambda_{\rm SW}\,. 
\end{equation} 
Here $A_i$, $B_i$ form a symplectic basis of (compact) one-cycles. In the classical limit the $a_i$'s can be interpreted as the eigenvalues of the expectation value of the adjoint field $\langle\phi\rangle$ forming part of the $\Ntwo$ vector multiplet. We find therefore 
\begin{equation}\label{classical_sw_diff}
\lambda_{{\rm SW},\,cl.} = v \tr \left( \frac{dv}{v-\phi} \right) ~,
\end{equation} 
since this differential has poles with residue $a_i$ at $z=a_i$ and therefore is the classical limit of the Seiberg--Witten differential. This suggests to interpret 
\begin{equation}\label{dt_over_t}
\left\langle  \tr\left( \frac{dv}{v-\phi} \right) \right\rangle = T(v)\,dv =  \frac{dt}{t} ~.
\end{equation} 
in the full quantum theory. That the differential $T$ is given by \erf{dt_over_t} has been already obtained in \cite{deboer}. We just comment that this differential is already well defined in the $\Ntwo$ theory and that it has the simple relation to the Seiberg--Witten differential $v \,T(v)\,dv = \lambda_{\rm SW}$.

Now we want to break to $\None$ by adding a tree level superpotential for the adjoint chiral multiplet 
\be\label{deltaW}
W(\phi) =  \sum_{k=0}^m \frac{g_k}{k+1} \tr (\phi^{k+1}) \,.
\ee
In the type IIA brane configuration this is realized by taking $m$ NS-fivebranes instead of only one on the right side (NS' branes). We introduce the complex coordinate \linebreak$w=x^7 + i x^8$. The worldvolume of the NS'-fivebranes have to span the lines $w = v-\varphi_i$ where $\varphi_i$ are the roots of $W'_{\mathrm{tree}}(x)= \sum_{k=0}^m g_k x^k = g_m \prod_{i=1}^m (x - \varphi_i) = 0$. In the following we will set the highest coupling $g_m$ ---\,which determines the degree of the superpotential\,--- to $g_m=1$. 

The M-theory lift of $\None$ brane configuration has been obtained in \cite{HOO,wittenii,boeroz}. In the M-theory fivebrane realization, large $v$ values correspond to semiclassical physics. In order to implement the superpotential \erf{Wtree} in the M-theory fivebrane configuration we have to impose the asymptotic boundary conditions \cite{boeroz}
\be\label{asympbc}
v\rightarrow \infty \; \left\lbrace \begin{array}{lll}
w \rightarrow 0	     &,& t \sim v^N \\
w \rightarrow W'(v)  &,& t \sim 4 \Lambda^{2N-N_f} v^{N_f-N}
\end{array}\right. 
\ee 
From these boundary conditions a holomorphic function $w(t,v)$ on the Riemann surface \erf{swcurve1st} can be constructed. The two assumptions that allow to find $w$ are that the $\Ntwo$ Seiberg--Witten curve remains unchanged\,\footnote{This fact can be seen in a na\"{\i}ve way. Breaking to $\None$ means we rotate the NS'-branes in the $(w,v)$-plane. However, looking from the $(t,v)$-plane, which is how the construction in \cite{edi} is regarded, one sees no change in the type IIA picture. In this way, one effectively sees the curve does not change.} and that $w(t,v)$ is a rational function with no additional poles at finite values of $v$.  These allow to constrain $w$ to the form
\be\label{wonswcurve}
w = N(v) + \frac{H(v)}{S(v)} \Big(t-P_N(v) \Big) \,.
\ee
We assumed here that potentially present double points in the Seiberg--Witten curve \erf{swcurve1st} are collected in the polynomial $S(v)$
 \begin{equation}
P_N^2 -4\Lambda^{2N-N_f}\prod_{f=1}^{N_f} (v+ m_f) = S_{N-n}(v)^2 \, Q_{2n}(v)\,,
\end{equation}  
where $Q(v)$ is a polynomial of degree $2n$ and $S(v)$ a polynomial of degree $N-n$. 
Let us momentarily denote the two branches of $t$ as $t_\pm = P_N\pm \sqrt{P^2_N-4\Lambda^{2N-N_f}\prod_{f=1}^{N_f} (v+ m_f) }$. We find then
\begin{equation}
w_\pm(t_\pm,v) = N(v) \pm H(v) \sqrt{Q(v)}\,.
\end{equation} 
The asymptotic boundary conditions fix $N(v)$ to be
\be\label{asymptotic_bcs}
N(v) = [H(v) \sqrt{Q(v)}]_+ = W'(v)\,,
\ee
and finally one finds that $w$ fulfills a quadratic equation
\be
w^2 - 2 W'_m(v) w + f_{m-1}(v) =0 \,.
\ee
where $f_{m-1}(v)$ is a polynomial of degree $m-1$ which can also be written as $f_{m-1}=N(v)^2- H(v)^2 Q(v)$. 

{\bf Note.} Of course there is nothing new in the above discussion except for one minor point: contrary to the original work in \cite{boeroz} we do not assume that the degree $m+1$ of the superpotential $W(v)$  is limited by $m<N$. We also allowed in the above discussion arbitrarily high values of $m$, and as one can convince oneself easily nothing changes in the discussion compared to \cite{boeroz}.  To limit oneself to add superpotentials of degree up to $m<N$ is of course motivated by the fact that in field theory operators of the form $\tr(\phi^{N+k})$ are not independent from the operators $\tr(\phi^k)$ with $k\in 1\ldots N$. They can be expressed as products of the latter ones in the form of multi-trace operators. Quantum mechanically these expressions get modified, and we will consider the higher power single trace operators as basic objects in the theory. This is also the standard way of treating these operators in Dijkgraaf--Vafa theory. \footnote{See however \cite{vijay} for the case of multi-trace operators.}

The $\None$ supersymmetric gauge theory with $N_f$ hypermultiplets in the fundamental representation and superpotential $W_{\rm tree}(\phi)$ is therefore described in M-theory by a Riemann surface. It is embedded in a three-dimensional complex space labelled by $v,t,w$ and defined by the two equations 
\begin{eqnarray}
\label{swcurve}
t^2 - 2 P_N(v) t + 4 \Lambda^{2N-N_f} \prod_{f=1}^{N_f} (v+m_f) &=&0\,,\\
\label{dvcurve} w^2 - 2 W'_m(v) w + f_{m-1}(v) &=&0 \,.
\end{eqnarray}
Furthermore the asymptotic boundary conditions impose the factorization conditions
\begin{eqnarray}
\label{swfactorization}
P_N(v)^2 - 4 \Lambda^{2N-N_f} \prod_{f=1}^{N_f}(v+m_f) = S_{N-n}(v)^2 \, Q_{2n}(v) \,,\\
\label{dvfactorization} W'_m(v)^2 - f_{m-1}(v) = H_{m-n}(v)^2 \, Q_{2n}(v) \,.
\end{eqnarray} 
Equation \erf{dvcurve} has precisely the form of the matrix model curve in Dijkgraaf--Vafa theory, whereas \erf{swcurve} has the form of Seiberg--Witten curve as said previously. The gauge theory and matrix model resolvents are given by 
\begin{eqnarray}
\left\langle \tr \left( \frac{dv}{v-\phi}\right) \right\rangle &=& T(v)\,dv = \frac{dt}{t} \,,\\
\left\langle  -\frac{1}{16\pi^2} \tr \left( \frac{\cW_ \alpha \cW^\alpha}{v-\phi}\right)  dv\right\rangle  &=& 2 R(v)\,dv = w dv\,.
\end{eqnarray}
The effect of the factorizations is to modify the Riemann surface one has in the $\Ntwo$ theory. After adding the tree level superpotential, the moduli space is lifted and we are restricted to some singular submanifolds in it. The remaining cycles of the Riemann surface are given by the single roots associated to the $Q_{2n}(v)$ polynomial.

\subsection{Suggestion of the duality}
\label{ssec:suggestion}

As seen, in the above construction we have obtained the two equations \erf{swcurve} and \erf{dvcurve} defining our M-theory fivebrane. We got also two factorizations for them, \erf{swfactorization} and \erf{dvfactorization}, which tell us where the singular points for that Riemann surface are.
These two curves, \erf{swcurve} and \erf{dvcurve}, have been traditionally regarded as the Seiberg--Witten and Dijkgraaf--Vafa curves, respectively. This is motivated by the fact that one {\it first} constructs the $\Ntwo$ curve, i.e. the SW curve, and {\it afterwards} performs the brane rotations to get $\None$, i.e. the DV curve \cite{DVpapers}. 
Let us go back for a moment to equation \erf{wonswcurve} and take into account the asymptotic
boundary contition \erf{asympbc}, which is $w = W_m'(v) + \frac{H(v)}{S(v)}(t - P_N(v))$.
This can be trivially inverted to find $t$ as a meromorphic function on the Riemann surface \erf{dvcurve}
\begin{equation}
t = P_N(v) + \frac{S(v)}{H(v)} \Big( w-W'_m(v) \Big) ~.
\end{equation} 
In fact $t$ regarded as a function on the Riemann surface \erf{dvcurve} has precisely the same properties as $w$ as a function on \erf{swcurve}: it is rational and does not have any poles at finite values of $v$. Note that this is the decisive condition which uniquely fixes the form of the Riemann surface \erf{dvcurve} once the asymptotic boundary conditions are supplied. But it also automatically gives $t$ as a rational function on this Riemann surface with the asymptotic behavior
\begin{equation}
\label{asymptotic_t}
v\rightarrow \infty \; \left\lbrace \begin{array}{lll}
t \rightarrow 0	     &,& w \sim v^m \\
t \rightarrow P_N(v)  &,& w \sim 0 \,.\end{array} \right.
\end{equation} 
These boundary conditions are consistent with the ones in \erf{asympbc} as long as \mbox{$N_f < N$.}\linebreak This implies that the factorization conditions \erf{swfactorization} and \erf{dvfactorization} could have been obtained by starting with the curve \erf{dvcurve} and then construct $t$ as a rational function without poles at finite values of $v$ on it and demanding the boundary conditions \erf{asymptotic_t}. In such an approach \erf{dvcurve} would be interpreted as the Seiberg--Witten curve of a gauge theory with $U(m)$ gauge group and $\Nt_f$ fundamental flavors whose masses are given by writing
\begin{equation}
f_{m-1} = 4\widetilde\Lambda^{2m-\Nt_f} (v + \widetilde m_f) \,,
\end{equation}  
where the number of flavors is at most $\Nt_f = m-1$. We see that there is a second ``dual" $\None$ supersymmetric gauge theory which we can associate with the system of algebraic curves \erf{swcurve}, \erf{dvcurve} thus giving rise to two ``dual" interpretations. Let us list the most striking features of this duality:

\begin{enumerate}
\item We are interchanging the two polynomials $P_N$ and $W'_m$, so we have a duality between two supersymmetric gauge theories with different gauge groups and matter content. One is a theory with gauge group $G=U(N)$ and $N_f$ matter hypermultiplets, whereas the other is a theory with gauge group $U(m)$ and up to $\Nt_f = m-1$ hypermultiplets. Of course, this implies that a meaningful dual interpretation as gauge dynamics exists only for $m>1$, i.e. we need at least cubic superpotentials!
\item In principle we can put as many NS' branes as we want. We might write the number of NS' branes as $m=N-k$, where $k$ ranges from all negative integers up to $N-1$ (since at least we would like to have one NS' fivebrane.) Now, the highest term in the superpotential goes like $\phi^{m+1}$; therefore if we do not want powers in the superpotential higher than $N$ ---\,the rank of the gauge group\,--- we must keep $k>0$. However, with this values set, in the dual interpretation we get
\beann
{\rm rank~} \wtil{G} &=& m \\
\deg \wtil{W} &=& m+k+1 \quad\mbox{(remember $k>0$ here.)}
\eeann
and thus we {\it necessarily} generate higher order terms for the dual superpotential.

One could do it all the way around, with $k\leq 0$. Then one has higher order terms in the traditional theory and lower ones in the dual.
\item The argument above supports the idea of {\bf duality} in this construction. If in one picture we have lower order operators, in the other we irremediably have higher order terms, and vice versa. Thus, applying twice the map we come back to our original theory.
\begin{figure}[htbp]
\begin{center}\label{fig.IIA}
\scalebox{.45}{
\input{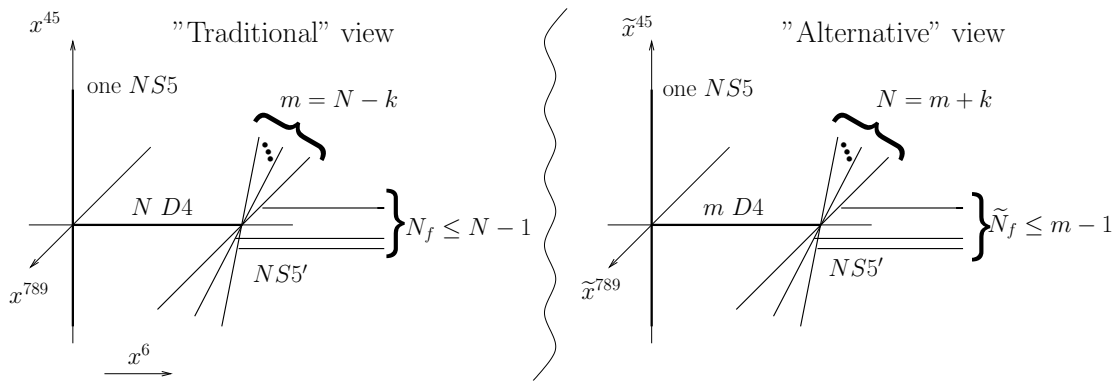}
 }
 \caption{\small \it A figure with the two possible (dual) descriptions of the same $\None$ equations 
 defining the M-theory fivebrane.}
\end{center}
\end{figure}
\item Note that one always has a {\it constraint} in the matter content of the form $N_f \leq N-1$ if the duality is to exist. This comes from consistency of the asymptotic boundary conditions in the dual interpretations or more simply from the fact that in the matrix model curve the degree of $f_{m-1}$ is determined through the condition that it represents normalizable or at most log-normalizable deformations! Therefore the constraint of allowing at most log-normalizable deformations is mapped through our duality to $N_f \leq N-1$. As is well known, gauge theories with $N_f\geq N$ are indeed qualitatively different from the one with $N_f<N$, e.g. complete higgsing of the gauge group is possible, baryonic operators can be formed and in the massless case moduli spaces exist instead of runaway vacua.
\item The relevant differentials entering the computation of gaugino condensates and ranks of the factor groups are
\begin{eqnarray}\label{dual_differentials_i}
T(v)\,dv = \frac{dt}{t} & 2 R(v) = w\,dv & \mbox{for the $U(N)$ gauge theory}\\
\widetilde T(v)\,dv =  \frac{dw}{w} & 2\Rt(v) = t\,dv & \mbox{for the $U(m)$ gauge theory} ~.\label{dual_differentials_ii} 
\end{eqnarray} 

\item Notice that the M-theory fivebrane construction does not physically give rise to this duality. In the M-theory setup the coordinate $t$ is singled out by the holomorphic threeform $\Omega = \frac{dt}{t}\wedge dw \wedge dv$, which determines the volume form $\Omega \wedge \overline\Omega$ and determines $\Im (\log(t))$ to be compact. The dual theory would demand the threeform to be $\widetilde \Omega = dt \wedge \frac{dw}{w} \wedge dv$ and therefore indicating that $\Im(\log(w))$ is compact. Of course the original M-theory/type IIA background is not compact in this direction. It has to be emphasized, however, that this information is not needed to compute the effective superpotential. All one needs is the embedding \erf{swcurve}, \erf{dvcurve} of the Riemann surface in the complex three dimensional space parameterized by $t,w,v$ and the differentials \erf{dual_differentials_i}, \erf{dual_differentials_ii}. In fact, as explicitly demonstrated in \cite{deboer}, the M-theory fivebrane does not reproduce the gauge theory physics correctly beyond that, e.g. K\"ahler terms computed from the M-theory fivebrane do not allow a gauge theory interpretation. So there is no reason to associate the M-fivebrane with the gauge theory beyond the information that can be obtained from the embedding of the Riemann surface in the three dimensional complex space and some choice of differentials $T$ and $R$ or $\widetilde T$ and $\Rt$. One might therefore interpret the space spanned by $t,v,w$ as an abstract space and forget about the M-theory origin of the construction. 
\end{enumerate}

\setcounter{equation}{0}
\section{Exact ``dual" superpotentials} \label{sec:dual-spots}

We want to compute explicitly now in some examples the exact superpotentials of the ``dual" gauge theories. We will do this in a strong coupling approach, viewing the tree level superpotential for the adjoint scalar as a perturbation of the $\Ntwo$ theory. We follow the work in \cite{CachaVafa}, where the equivalence of the superpotential computed from the geometry with the one computed from a low energy field theoretical approach has been proved for the $U(N)$ super Yang--Mills case. This has been extended later in \cite{Ookuchi} to the case with flavors with  $N_f < N$ and in \cite{AFOS} for the case with $N\leq N_f<2N$.

However both \cite{Ookuchi} and \cite{AFOS} considered superpotentials of at most degree $N$, i.e. the highest adding terms of the form $\tr (\phi^k)$ with $k_\mathrm{max} = m +1 \leq N$. As we have seen before, our duality forces us to consider the cases with $k_{\rm max} > N$ as well, because if the original theory has a superpotential of degree less or equal $N$, the dual will be a theory with a superpotential of degree $\widetilde m +1 \leq N+1$ and gauge group $U(\Nt)$, where $\Nt = m$. The operators with $k> N$ (or $\widetilde k > \Nt$) are of course not independent from the ones with lower powers and can be expressed as function of the latter. This has to be imposed as additional constraints on the low energy superpotential as in \cite{CSWnomatter}.

We therefore want to prove in the following the equivalence of the low energy field theory superpotential, denoted by $W_{\rm low}$ with the superpotential computed from geometry denoted by $W_{\rm eff}$, in the case with general superpotential and $N_f < 2N$.\,\footnote{Even though for the duality to hold we will always have $N_f < N$, the rank of the gauge group.}  Following \cite{CachaVafa}, we have to show
\bea \label{eq:CVconds1}
W_{\rm low}(g_k,\Lambda,m_f) \Big|_{\Lambda\to 0} &=& W_{\rm eff}(\vev{S_i}) \Big|_{\Lambda\to 0} ~, \\ \label{eq:CVconds2}
\frac{\partial W_{\rm low}(g_k,\Lambda,m_f)}{\partial\log\Lambda^{b_0}} &=& \frac{\partial W_{\rm eff}(\vev{S_i})}{\partial\log\Lambda^{b_0}} ~,
\eea
where $b_0 = 2N-N_f$ is the one-loop beta-function coefficient for $U(N)$ gauge group.

\subsection{Field theory superpotential}
We first show that the matrix model curve follows from the field theory superpotential.

The moduli space for the Coulomb branch of the $\Ntwo$ theory is $N$ dimensional. Once we break to
$\None$ by adding the tree level superpotential \erf{Wtree}, its effect is to lift the moduli space except for some singular submanifolds, where $N-n$ massless dyons appear. We parameterize the moduli space by $U_k = \frac{1}{k}\langle\tr(\phi^k)\rangle$. 

The polynomial  entering the Seiberg--Witten curve is defined as
\[
P_N(v) = \langle\mathrm{det} (v-\phi)\rangle ~,
\] 
which using the solution for $T(v)$ can be written as \cite{CSWmatter}
\be \label{eq:P_Nquantum}
P_N(v) = v^N \exp\left( -\sum_{r=1}^\infty \frac{U_r}{v^r} \right) + \frac{F_{N_f}(v)}{4v^N} \exp\left( +\sum_{r=1}^\infty \frac{U_r}{v^r} \right) ~,
\ee
where we denoted $F_{N_f} = 4 \Lambda^{b_0} \prod_{f=1}^{N_f} (v+m_f)$. The negative powers in $v$ give the quantum corrected relations between the $U_k$ with $k>N$ and the independent single trace operators with $k\leq N$.

The low energy superpotential is given by \footnote{There is an implicit $(2\pi i)^{-1}$ in the definition of $\oint$.}
\bea
W_{\rm eff} &=& \sum_{k=0}^m g_k U_{k+1} + \oint V_{m-N}(x) \left( x^N e^{\textstyle -\sum_{r=1}^\infty \frac{U_r}{x^r}} + \frac{F_{N_f}(x)}{4x^N} \,e^{+\textstyle \sum_{r=1}^\infty \frac{U_r}{x^r}} \right) dx \nn \\ \label{eq:Weff-multipliers}
&+& \sum_{i=1}^{N-n} \left( L_i\oint \frac{P_N(x) + \epsilon_i\sqrt{F_{N_f}(x)}}{x-p_i} dx + B_i\oint \frac{P_N(x) + \epsilon_i\sqrt{F_{N_f}(x)}}{(x-p_i)^2} dx \right) ~. \hspace*{1cm}
\eea
$V_{m-N}(x)$ is a polynomial of order $m-N$ whose coefficients are the Lagrange multipliers enforcing the relations for the higher power $U_k$'s. On the singular submanifold the Seiberg--Witten curve has $N-n$ double points $p_i$. These double points are enforced by the Lagrange multipliers $L_i$ and $B_i$, furthermore  $\epsilon_i =  \pm 1$. The equations of motion for $B_i$ and $p_i$ tell us that, if we do not want to have triple or higher roots, then $B_i=0$ on-shell. Differentiating with respect to $U_{r+1}$ yields
\be
\frac{\partial W_{\rm eff}}{\partial U_{r+1}} = g_r + \oint V_{m-N}(x)\frac{\partial P_N(x)}{\partial U_{r+1}} \,dx + \sum_{i=1}^{N-n} \oint \frac{L_i}{x-p_i} \frac{\partial P_N(x)}{\partial U_{r+1}} \,dx = 0 ~,
\ee
where using \erf{eq:P_Nquantum} we obtain
\be
\frac{\partial P_N(x)}{\partial U_{r+1}} = \frac{1}{x^{r+1}} \left[ P_N(x) - 2x^N  \exp\left( -\sum_{i=1}^\infty \frac{U_i}{x^i} \right) \right]_+ = - \left[ \frac{\sqrt{P_N^2 - F_{N_f}}}{x^{r+1}} \right]_+ ~.
\ee

Thus we have an expression for the $g_k$'s, which we can multiply by $v^k$ and sum over $k$ to obtain the derivative of the tree level superpotential
\be \label{eq:W'(z)prvnal}
W'(v) = \sum_{k=0}^\infty \left( \oint V_{m-N}(x) + \sum_{i=1}^{N-n} \oint \frac{L_i}{x-p_i} \right) \frac{v^k}{x^{k+1}} \sqrt{P_N(x)^2 - F_{N_f}(x)} ~dx ~,
\ee
where we have extended the summation to $\infty$ since the terms for $k>m$ do not contribute. We write
\be
\sum_{i=1}^{N-n} \frac{L_i}{x-p_i}= \frac{{\widehat H}_{N-n-1}(x)}{S_{N-n}(x)} ~,
\ee
and use the factorization of the SW curve to obtain
\be
W'(v) = \oint \frac{y_{m.m.}(x)}{x-v}  ~dx ~,
\ee
where the integral is over a large contour at infinity and
\bea
y_{m.m.}^2 &=& H^2_{m-n}(v)\,Q_{2n}(v) = W'_m(v)^2 + \cO(v^{m-1}) ~, \\
{\rm with}~~ H_{m-n} &=& V_{m-N}S_{N-n} + {\widehat H}_{N-n-1} ~. \nn
\eea
This gives a strong coupling field theory derivation of the matrix model curve in the case valid for $N_f < 2N$ and arbitrary degree superpotential.

Since the breaking pattern is $U(N) \to \prod_{i=1}^n U(N_i)$, the classical limit of the superpotential
is 
\be
W_{\rm low, classical} = \sum_{i=1}^n N_i W(\varphi_i) \,.
\ee

We now compute the dependence of the low energy superpotential on the scale $\Lambda^{b_0}$:
\bea
\frac{\partial W_{\rm low}(g_k,\Lambda)}{\partial\log\Lambda^{b_0}} &=& \oint_{\cC_{\rm large}} V_{m-N}(x) \frac{F_{N_f}(x)}{4x^N} ~e^{\textstyle +\sum_{r=1}^\infty \frac{U_r}{x^r}} dx + \sum_{i=1}^{N-n} \frac{L_i\,\epsilon_i}{2} \sqrt{F_{N_f}(p_i)} \nn \\
\label{eq:dWlow/dlog} &+& \oint_{\cC_{\rm large}} \sum_{i=1}^{N-n} \frac{L_i}{x-p_i} \frac{F_{N_f}(x)}{4x^N} e^{\textstyle +\sum_{r=1}^\infty \frac{U_r}{x^r}} dx ~. \qquad
\eea

From the geometry it follows (as we will review shortly in the next section) that the logarithmic derivative with respect to $\Lambda^{b_0}$ is given by the highest coefficient $b_{m-1}$ of the polynomial $f_{m-1}= \sum_{i=0}^{m-1} b_i\,v^i$.

In order to compute $b_{m-1}$ we note that \footnote{Remember that we have set the coupling $g_m=1$.}
\be \label{eq:massage}
W'_m(v) - \sqrt{W'_m(v)^2 - f_{m-1}(v)} = \frac{b_{m-1}}{2}\,\frac{1}{v} + \cO\Big( \frac{1}{v^2} \Big) ~.
\ee

Using both the factorizations of the Dijkgraaf--Vafa and the Seiberg--Witten curves, we can write
\bea \label{eq:massage1}
\sqrt{W'_m(v)^2 - f_{m-1}(v)} &=& H_{m-n} \sqrt{Q_{2n}} = (V_{m-N}S_{N-n} + \widehat H_{N-n-1}) \sqrt{Q_{2n}} = \nn \\
&=&\left( V_{m-N}(v)+\sum_{i=1}^{N-n} \frac{L_i}{v-p_i} \right) \sqrt{P_N(v)^2-F_{N_f}(v)} ~.\qquad
\eea
As we have just seen $W'(v)$ can be written as
\be
W'(v) = \oint_{\cC_{\rm large}} \left( V_{m-N}(x)+\sum_{i=1}^{N-n} \frac{L_i}{x-p_i} \right) \frac{ \sqrt{P_N(x)^2-F_{N_f}(x)} }{x-v} \,dx ~.
\ee

\begin{figure}[htbp]
\begin{center}
\scalebox{.65}{
\input{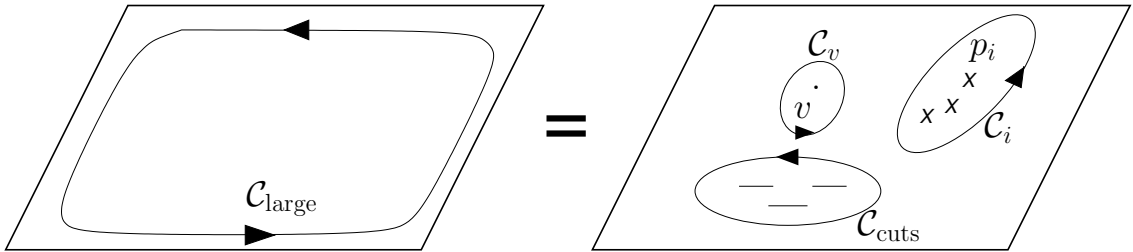}
 }
 \caption{\label{fig:contours}\small\it The contour $\cC_{\rm large}$ can be deformed into a contour centered at $v$ and
 a contour $\cC_{\rm cuts}$ around the cuts of the Riemann surface. Since the integrand is regular at
 $p_i$ the contour $\cC_i$ does not contribute.}
\end{center}
\end{figure}

We can deform the large contour $\cC_{\rm large}$ into a contour $\cC_v$, encircling just the point $x=v$ and $\cC_{\rm cuts}$, which encloses the branch cuts of the Riemann surface (see fig. \ref{fig:contours}). Notice that the integrand is regular at $x=p_i$ due to the presence of the double points inside the square root. The contour surrounding $x=v$ yields the same expression as \erf{eq:massage1}, so using equation \erf{eq:massage} we find
\bea
&& \hspace*{-18mm} \frac{b_{m-1}}{2}\,\frac{1}{v} + \cO(v^{-2}) = \oint_{\cC_{\rm cuts}} \left( V_{m-N}(x)+\sum_{i=1}^{N-n} \frac{L_i}{x-p_i} \right) \frac{ \sqrt{P_N(x)^2-F_{N_f}(x)} }{x-v} \,dx \\
&=& - \frac{1}{v} \oint_{\cC_{\rm cuts}} \left( V_{m-N}(x)+\sum_{i=1}^{N-n} \frac{L_i}{x-p_i} \right) \sqrt{P_N(x)^2-F_{N_f}(x)} \,dx + \cO(v^{-2}) ~, \quad \label{one_over_v}
\eea
where in the second line we have considered $v\gg 1$. To match this with \erf{eq:dWlow/dlog} we still have to do some manipulations. We notice that
\be
\sqrt{P_N^2 - F_{N_f}} - P_N  = -\frac{F_{N_f}}{2x^N} \,e^{\textstyle +\sum\frac{U_r}{x^r}} ~. \qquad
\ee
For the first term inside the brackets of \erf{one_over_v} we note that we can subtract a $V_{m-N} P_N$ term from the integrand since it is analytic on the whole plane. Thus
\bea
&& \oint_{\cC_{\rm cuts}} V_{m-N} \sqrt{P_N^2 - F_{N_f}} dx = \oint_{\cC_{\rm cuts}} V_{m-N} \left( \sqrt{P_N^2 - F_{N_f}} -P_N \right) dx = \nn \\
&=& \oint_{\cC_{\rm large}} V_{m-N} \left( \sqrt{P_N^2 - F_{N_f}} -P_N \right) dx = - \oint_{\cC_{\rm large}} V_{m-N}\,\frac{F_{N_f}}{2x^N} \,e^{\textstyle +\sum\frac{U_r}{x^r}} dx ~, \qquad\qquad
\eea
where the last contour deformation could be performed without any problem.

Now we consider the second term in the brackets of \erf{one_over_v}. 
\bea
&& \oint_{\cC_{\rm cuts}} \sum_{i=1}^{N-n} \frac{L_i}{x-p_i} \sqrt{P_N^2 - F_{N_f}} \,dx = \oint_{\cC_{\rm cuts}} \sum_{i=1}^{N-n} \frac{L_i}{x-p_i} \left( \sqrt{P_N^2 - F_{N_f}} -P_N \right) \,dx \nn \\
&=& \left( \oint_{\cC_{\rm large}} - \oint_{\cC_i} \right) \sum_{i=1}^{N-n} \frac{L_i}{x-p_i} \left( \sqrt{P_N^2 - F_{N_f}} -P_N \right) \,dx  \\
&=& -\oint_{\cC_{\rm large}} \sum_{i=1}^{N-n} \frac{L_i}{x-p_i}\,\frac{F_{N_f}}{2x^N} \,e^{\textstyle +\sum\frac{U_r}{x^r}} dx - \oint_{\cC_i} \sum_{i=1}^{N-n} \frac{L_i}{x-p_i} \left( \sqrt{P_N^2 - F_{N_f}} -P_N \right) \,dx \nn \\
&=& -\oint_{\cC_{\rm large}} \sum_{i=1}^{N-n} \frac{L_i}{x-p_i}\,\frac{F_{N_f}}{2x^N} \,e^{\textstyle +\sum\frac{U_r}{x^r}} dx - \sum_{i=1}^{N-n} L_i \left( \sqrt{P_N(p_i)^2 - F_{N_f}(p_i)} -P_N(p_i) \right) ~. \nn
\eea
For the last term, we note that $\sqrt{P_N(p_i)^2-F_{N_f}(p_i)}=0$, and also $P_N(p_i) = -\epsilon_i \sqrt{F_{N_f}(p_i)}$. Therefore this takes us to the end of the computation and we find
\bea
\frac{b_{m-1}}{2} &=& \oint_{\cC_{\rm large}} V_{m-N}\,\frac{F_{N_f}}{2x^N} \,e^{\textstyle +\sum\frac{U_r}{x^r}} \, dx + \oint_{\cC_{\rm large}} \sum_{i=1}^{N-n} \frac{L_i}{x-p_i} \,\frac{F_{N_f}}{2x^N} \,e^{\textstyle +\sum\frac{U_r}{x^r}} dx \nn \\
&+& \sum_{i=1}^{N-n} L_i \epsilon_i \sqrt{F_{N_f}(p_i)} ~.
\eea

Comparison with \erf{eq:dWlow/dlog} shows that
\be \label{eq:dWlow/b_m-1}
\frac{\partial W_{\rm low}(g_k,\Lambda)}{\partial\log\Lambda^{b_0}} = \frac{b_{m-1}}{4g_m} ~,
\ee
for a number of flavors $N_f < 2N$ and arbitrary degree of the tree level superpotential.
For the sake of completeness we also have reinstated the coupling $g_m$.

\subsection{``Dual'' geometric superpotential}
In the preceding subsection we have identified the expressions for $W_{\rm low}$ in the classical limit and its derivative with respect to $\log\Lambda^{b_0}$. Now we do the same using the Riemann surface. If at low energy we have $U(1)^n$ ``photons'', some of our branch cuts will be closed. Since the degree of the superpotential is $m\geq N$, $m-n$ cuts will be closed. From \cite{AFOS} we should assume some $S_i=0$ from the beginning, implemented by hand off-shell.\,\footnote{The question when and if $S_i=0$ implies a closed cut has been thoroughly investigated in \cite{IKRSV}. It turned out to be especially important in the case of theories with symplectic or orthogonal gauge group \cite{kraus,alday,cachazo,matone} where non vanishing cuts for $S_i=0$ could be traced back to non-trivial effects of orientifolds in the string theory realizations \cite{LL}.} We will choose the first $n$ $S_i$'s non-vanishing, and the remaining $m-n$ as zero. From this we can infer the ``dual'' effective superpotential following \cite{CSWmatter}. For pseudo-confining vacua, and in the classical limit, we have
\bea
W_{\rm eff} &=& -{1\over 2}\sum_{i=1}^n N_i \int_{\hb_i^r} y_{m.m.}(z)dz - {1\over 2}\sum_{f=1}^{N_f} \int_{-m_f}^{\widetilde\Lambda_0} y_{m.m.}(z)dz \nn \\
&+& {1\over 2}(2N-N_f) W(\Lambda_0) + {1\over 2} \sum_{f=1}^{N_f} W(-m_f) \nn \\
&-& \pi i (2N-N_f)S + S \log\left( \frac{ \Lambda^{2N-N_f}}{\Lambda_0^{2N-N_f}} \right) + 2\pi i \sum_{i=1}^{n-1} c_i S_i ~.
\eea
where $\hb_i$ are compact cycles that together with the first $n-1$ $A_i$ cycles form a symplectic basis, $c_i$ are the (integer-) periods of the $T$ differential on these cycles and $\Lambda_0$ is a cut-off. The classical limit has been computed in \cite{CSWmatter} and in the pseudo-confining branch it was shown to be
\be \label{eq:Weff-classical}
W_{\rm eff}\big|_{\Lambda\to 0} = \sum_{i=1}^n N_i \, W(\varphi_i) ~.
\ee
The derivative of the effective superpotential with respect to the logarithm of the scale is
\be \label{eq:b_m-1strings}
\frac{\partial W_{\rm eff }}{\partial \log\Lambda^{b_0}} = \sum_{i=1}^m S_i = S = \frac{b_{m-1}}{4 g_m} ~.
\ee
We used the fact that $S$ can be computed easily from the matrix model curve as a period integral
over a large circle of $S =\frac 1 2 \oint_{\cC_{\rm large}} y_{m.m.}(z)\,dz$ \cite{CachaVafa}.
Hence we can obtain the value of the derivative
\be \label{eq:dWeff/b_m-1}
\frac{\partial W_{\rm eff}(\vev{S_i},\Lambda)}{\partial\log\Lambda^{b_0}} = S =  {b_{m-1} \over 4 g_m} ~,
\ee
where in the last identity we used \erf{eq:b_m-1strings}.

This shows the equivalence between the field theory and the ``dual'' geometry effective superpotentials. The argument as it has been presented is valid only for the pseudo-confining vacua.
In the quantum theory the Higgs vacua are continuously connected to them by varying the masses $m_f$ in such a way that they cross a branch cut of the matrix model curve. We can compute the superpotential in an arbitrary vacuum by first computing it on the pseudo-confining branch and the analytically continuing the values of $m_f$. In this way the equivalence extends to the pseudo-higgs branches as well.

\subsection{Examples} \label{ssc:examples}
In this subsection we discuss a couple of examples to show how the duality claimed in subsection \ref{ssec:suggestion} works. We will consider first the example of a $U(2)$ theory with no flavors and cubic superpotential. We trivially find the same Riemann surface for both descriptions and thus the dual theory is the same as the original one.
 
As a second example we will consider the case of a $U(3)$ gauge group without fundamental matter and again a cubic tree level superpotential. This is how we regard the traditional picture of the M-theory curve, therefore obtaining the respective results in the dual interpretation.

$P_N(v)$ is given in terms of the symmetric polynomials $s_i$ as
\be
P_N(v) = \det(v-\phi) = v^N + s_1 v^{N-1} + \ldots + s_N ~,
\ee
The factorization conditions determine the vacuum expectation values of the $s_i$'s and from \erf{eq:P_Nquantum} we find the vevs of the Casimirs $U_k$. These values have to be simply plugged into the tree-level superpotential to compute $W_{\mathrm low}$.

\noindent{\textit{\bfseries\boldmath `\,$U(2)$ gauge theory, no matter'}}\\[2ex]
Starting with the traditional picture (this is a choice), we have $N=2$, $N_f=0$ and $\deg W'=2$. Since $P_N = v^2 + s_1 v + s_2$ , the SW curve is
\be
{\rm SW:} \quad (v^2+s_1 v +s_2)^2 - 4\Lambda^4 ~,
\ee
whereas we take $W_{\rm tree}$ as
\be
W_{\rm tree} = U_3 + mU_2 + \ell U_1 \quad \to \quad W_{\rm tree}(v) = \frac{1}{3}v^3+ \frac{m}{2}v^2 + \ell v ~,
\ee
so the DV curve is
\be
{\rm DV:} \quad (v^2+ mv + \ell)^2 - (f_1 v +f_0) ~.
\ee
The idea now is to solve for $\{s_1,s_2,f_0,f_1\}$ in terms of the quantities of the theory by using factorization of the curves. The trivial case corresponds to the choice $n=2$ (could be $n=1$ too.) Then both $S(v)$ and $H(v)$ are constants which we can set to be equal. Hence, the two curves are automatically equal so
\be
(v^2+s_1 v +s_2)^2 - 4\Lambda^4 = (v^2+ mv + \ell)^2 - (f_1 v +f_0) ~,
\ee
from where one obtains
\be \label{eq:quants1}
s_1 = m ~~,~~ s_2 = \ell ~~,~~ f_0 = 4\Lambda^4 ~~,~~ f_1 = 0 ~.
\ee
Now, for this theory the $U_k$'s are given by
\be \label{eq:Ukof(sk)1}
U_1 = -s_1 ~,\quad \displaystyle U_2 = \frac{U_1^2}{2}-s_2 ~,\quad \displaystyle U_3 = U_1\,U_2 - \frac{U_1^3}{6} ~,
\ee
so once we found the values for $s_1,s_2$ we use \erf{eq:Ukof(sk)1} and \erf{eq:quants1} to find
\begin{eqnarray}
&& U_1 = -m ~~,~~ U_2 = \frac{m^2}{2}-\ell ~~,~~ U_3 = m\ell - \frac{m^3}{3} ~, \\
&& W_{\rm low} = \frac{m^3}{6}-m\ell ~.
\end{eqnarray}

As said, this is just one way of regarding the projections of the M-theory curve. The alternative (dual) one is to do the change $\rm SW \to \widetilde{DV}$ and $\rm DV \to \widetilde{SW}$. However, in this trivial example the new theory we obtain is also a $U(2)$ gauge theory without matter (since $f_1=0$). Because the factorization problem is the same (this is always true), the low energy effective superpotential has the same shape in this picture
\be
W_{\rm low}^{\,dual} = \frac{m^3}{6}-m\ell ~.
\ee

\noindent{\textit{\bfseries\boldmath `\,$U(3)$ gauge theory, no matter'}}\\[2ex]
Here we consider a somewhat more interesting case. We will regard the traditional picture as that with $N=3$, $N_f=0$ and $\deg W'=2$. The curves are
\bea
{\rm SW:} && (v^3 + s_1 v^2 + s_2 v + s_3)^2 - 4\Lambda^6 ~, \\
{\rm DV:} && (v^2 + m v + \ell)^2 - (f_1 v + f_0) ~.
\eea

For the choice of $n=2$, $H(v)$ is a constant. The factorization problem thus translates into
\be
(v^3 + s_1 v^2 + s_2 v + s_3)^2 - 4\Lambda^6 = (v-a)^2 \left[ (v^2 + m v + \ell)^2 - (f_1 v + f_0) \right] ~,
\ee
having four different solutions. Defining $\Delta^2 \equiv m^2 -4\ell$, they are reached through
\[ \Lambda^3 \to -\Lambda^3 ~, \quad \Delta \to -\Delta ~. \]
To be concise we will consider the solution with plus signs for $\Lambda^3$ and $\Delta$
\be
\ba{lclcl}
\displaystyle s_1 = \frac{3m-\Delta}{2} &~,~~& \displaystyle s_2 = \ell + \frac{m(m-\Delta)}{2} &~,~~& \displaystyle s_3 = \frac{\ell(m-\Delta)}{2} +2\Lambda^3 \\ \\
f_0 = -2 \Lambda^3 (m+\Delta) &~,~~ & f_1 = -4\Lambda^3  &~,~~& \displaystyle a = \frac{\Delta-m}{2} ~.
\ea
\ee
From these one can obtain the $U_k$'s
\be
U_1 = \frac{\Delta-3m}{2} ~,\quad U_2 = \frac{m(3m-\Delta)}{4} - \frac{3\ell}{2} ~,\quad U_3 = \frac{3m\ell}{2} + \frac{\Delta(m^2-\ell)}{6} - \frac{m^3}{2} -2\Lambda^3 ~,
\ee
which allows us to write the effective superpotential
\be
W_{\rm low} = \frac{m^3}{4} - \frac{3m\ell}{2} - \frac{\Delta^3}{12} - 2\Lambda^3 ~.
\ee

Let us now go to the dual picture and see what the superpotential is in terms of other variables. From the previous calculations we already know all of the quantities. The curves are
\bea
\widetilde{\rm SW}: && (v^2 + m v + \ell)^2 - (f_1 v + f_0) ~, \\
\widetilde{\rm DV}: && (v^3 + s_1 v^2 + s_2 v + s_3)^2 - 4\Lambda^6 ~.
\eea
so then we know
\be
\left\{ \ba{l} \st_1 = m ~\,,~~ \st_2 = \ell \\ \widetilde{F}_0 = f_0 ~,~~ \widetilde{F}_1 = f_1 \\ \mbox{Couplings of superpotential: } \{ s_1,s_2,s_3 \} \ea \right.
\ee
As one can see, for this dual theory we have matter and a {\it quartic} superpotential with different couplings, which we write as
\be
\Wt_{\rm tree} = \Ut_4 + s_1\Ut_3 + s_2\Ut_2 + s_3\Ut_1 ~.
\ee
There is one dual flavor with mass $\widetilde m = (m+\Delta)/2$ and the scale of the dual theory is $\widetilde\Lambda^3 = \Lambda^3$\,! Since in this case matter is present the relations between the $s_i$'s and the $U_k$'s suffer a modification 
\bea
&& \Ut_1 = -\st_1 ~,\quad \Ut_2 = \frac{\Ut_1^2}{2} - \st_2 ~,\quad \Ut_3 = \Ut_1\,\Ut_2 - \frac{\Ut_1^3}{6} + \frac{\widetilde{F}_1}{4}  \qquad \qquad \\
&& \Ut_4 = \Ut_1\,\Ut_3 + \frac{\Ut_2^2}{2} + \frac{\Ut_1^4}{24} - \frac{\Ut_1^2\,\Ut_2}{2} + \frac{\widetilde{F}_1\Ut_1}{4} + \frac{\widetilde{F}_0}{4} ~, \qquad
\eea
so plugging the expressions we know for $\{ \st_1,\st_2,\widetilde{F}_0,\widetilde{F}_1 \}$ yields
\bea
\Ut_1 = -m &,& \Ut_3 = m\ell - {m^3\over 3} + \Lambda^3 \\
\Ut_2 = {m^2\over 2}-\ell &,& \Ut_4 = {m^4\over 4} + {\ell^2\over 2} - {3m+\Delta \over 2}\Lambda^3 -m^2\ell ~,
\eea
while further plugging in the values of $\{s_1,s_2,s_3\}$ (which are interpreted as couplings now)
finally gives the effective superpotential for this dual interpretation
\be
\frac{1}{g_3} W_{\rm low}^{\,dual}={m\ell\Delta\over 6} - {m\Delta^3\over 12} - {\ell^2\over 2} -2m\Lambda^3  ~.
\ee
See the results obtained are different for these two theories, even though they come from the same M-theory curve.

\setcounter{equation}{0}
\section{Discussion and Conclusions} \label{sec:conclusions}

We have seen that the M-theory curve describing $\None$ gauge theories allows two dual interpretations upon exchanging what one views as the Seiberg--Witten or as the Dijkgraaf--Vafa curve. There are two gauge theories which give rise to the same curve embedded in a three dimensional complex space. The dual interpretations are possible only for the groundstate since the fluctuations in the two systems are different. However the factorization conditions contain enough information to compute the low energy effective superpotential in both theories. 

The duality exchanges the rank of the gauge group $N$ with the degree of the superpotential $m+1$ and furthermore the deformations of the Dijkgraaf--Vafa curve with the mass terms of the Seiberg--Witten curve. Because of the exchange of $m$ and $N$ one always has to deal with a tree level superpotential of high degree, i.e. operators $\tr(\phi)$ with $k>N$. We therefore had to extend the known proofs in the literature of the equivalence of the field theory effective superpotential with the DV (or matrix model) superpotential to cases of arbitrary $k$ and a number of flavors $N_f < 2N$. Another intriguing aspect of the duality is that the constraint of allowing at most log-normalizable deformations of the DV curve is mapped to a constraint in the number of flavors $N_f<N$. This is rather interesting since it is well known that the low energy physics of gauge theories with more flavors, although being qualitatively different, is still well defined. So one might ask then what is the significance of the deformations exceeding the log-normalizable bound in the DV curve. We leave this question for future investigation.

\bcen{\large\bf Acknowledgements}\ecen
The research of K. L. is supported by the Ministerio de Ciencia y Tecnolog\'{\i}a through a Ram\'on y Cajal contract and by the Plan Nacional de Altas Energ\'{\i}as FPA-2003-02-877. The research of S. M. is supported by an FPI 01/0728/2004 grant from Comunidad de Madrid and by the Plan Nacional de Altas Energ\'{\i}as FPA-2003-02-877. S. M. also wants to thank G. S\'anchez for her support.


\end{document}